# A Comparative Analysis of Institutional and Course Generative AI Policies within Higher Education: Implications for Instruction in Computing Education

Amrita Ganguly
George Mason University
Fairfax, VA, USA
agangul@gmu.edu

Aditya Johri
George Mason University
Fairfax, VA, USA
johri@gmu.edu

Nora McDonald
George Mason University
Fairfax, VA, USA
nmcdona4@gmu.edu

Areej Ali
George Mason University
Fairfax, VA, USA
aali40@gmu.edu

Umama Dewan
George Mason University
Fairfax, VA, USA
udewan@gmu.edu

Aayushi Hingle Collier
Montgomery College
Rockville, MD , USA
aayushi.hingle@montgomerycollege.edu

***Abstract*—With the increased use of generative AI (GenAI) applications such as ChatGPT, higher education institutions (HEIs) have released a range of guidelines and policies to direct adoption within their institutions. In computer science (CS) courses GenAI adoption is especially high and the implications for student learning are significant. At the same time, instructors have also been forced to address the use of GenAI as students have started to use it for a range of functions. Currently, comparative analysis of guidance provided by institutions and its uptake in instruction is lacking. In this paper we bridge this gap by comparing institutional and computing course level guidance to better understand this terrain. We utilize secondary analysis of institutional and course syllabi guidelines from higher education institutions in the U.S. classified as research-intensive. Our findings reveal that although institutional guidance is more pro-use, at the course-level the uptake is still guarded. We discuss the implications and propose an instructor-centered framework to guide future adoption of GenAI.**

## I. Introduction

The terrain of higher education is constantly in flux and GenAI, with the release of ChatGPT in late 2022, has created major ripples in the higher education landscape. GenAI's capabilities for generating human-like text, problem solving, and potential use as a tutor and coach, foreshadow big shifts for critical functions within higher education including teaching and learning, research, and administrative tasks. These changes can also be noticed in computing education. GenAI tools that can generate, explain, and debug code are challenging traditional ways of teaching and assessing programming and computational thinking. Furthermore, both the high and fast adoption of GenAI by students and other stakeholders has meant that many higher education institutions (HEIs) are playing catch up to the technology and increasingly concerned about its future impact.

While GenAI has shown some potential for positive impact, its use has come with concerns related to academic integrity, privacy and ethics, and lack of learning and metacognitive awareness among students (McDonald et al. 2025). For example, students may use GenAI to complete programming assignments without developing the underlying problem-solving skills that are central to computing education. Therefore, GenAI's use has created a paradox in that there are both potential opportunities but also significant challenges (An et al. 2025)(Jin et al. 2025). To minimize the challenges and adverse outcomes, universities have recognized the need to provide clear guidance for responsible use of GenAI that includes policies and regulations for different use cases and stakeholders (Moore and Lookadoo 2024).

A range of guidance has been issued at the institutional level (An et al. 2025; Jin et al. 2025; McDonald et al. 2025). However, as in other areas of AI, the large amount of changing guidance and rapid technological change makes it difficult to identify what is effective and actionable (Cacho 2024). As a result, although some institutional guidance appears in course-level policies, it is still unclear how these relate to higher-level guidance, and there is also inconsistency across levels, where institutional policies may restrict GenAI use but allow exceptions, or instructors may prohibit its use despite institutional encouragement (Xiao et al. 2023). Consequently, despite the increased guidance available, a gap remains in understanding how instructors translate and implement institutional guidelines into course syllabi (Lee et al. 2024). Understanding this is important to ensure responsible use of GenAI so that students do not lose important lifelong skills while supporting educational goals of technical skill development and knowledge building (Cacho 2024).

In this paper we present a study that addresses this gap through a comparative analysis of policies from the same set of institutions to identify the relationship between institutional and computing course guidelines. Based on our literature review, this is the first study of its kind that maps the relationship between GenAI guidance at different levels using data from the same institution. We then discuss implications of this analysis to identify existing challenges and potential steps that can be taken to mitigate negative consequences of adoption, especially in the long term.

## II. Relevant work

### A. Institutional GenAI guidance in HEIs

An et al. (2025) examined guidance at the top 50 U.S. universities using sentiment analysis and found generally positive attitudes toward GenAI across institution types. They also identified significant differences between faculty- and student-focused guidelines, with 94% of universities including faculty guidance emphasizing course-specific GenAI policies. Consistent with prior studies, academic integrity and privacy emerged as dominant themes. The study concludes that, given the rapid evolution of GenAI, higher education institutions should develop flexible, stakeholder-specific policies that balance both opportunities and challenges. Wang et al. (2024) analyzed GenAI policies and resources from the top 100 U.S. universities and found that most universities take an open but cautious approach. The main concerns were ethical use, output accuracy, and data privacy. Most universities also provided resources such as syllabus templates, workshops, articles, and consultations. A study of policies from 40 universities across six geographic areas similarly found that universities mainly focused on academic integrity, teaching and learning, and equity (Jin et al. 2025).

One relevant finding from these studies is that HEI guidelines and policies generally view GenAI use as a positive force for education. Christ-Brendemühl (2025) found that 56.7% of university guidelines state that the opportunities of GenAI outweigh the risks. In addition, 73.1% of institutions permit staff use of GenAI, and 83.6% allow student use. Overall, the main theme is acceptance and recognition of AI's potential to support learning. Another common finding, as shown by Moorhouse et al. (2023), is a strong focus on faculty responsibility in implementing GenAI in institutions. Faculty are advised to test assessment tasks with GenAI and to integrate student use of GenAI into assessment design. Similarly, Wang et al. (2024) highlight practical implications for educators, including accepting GenAI presence, aligning its use with learning objectives, adapting curriculum to reduce misuse, and using multiple evaluation strategies. Overall, much of the guidance places responsibility on faculty to actively respond to and manage GenAI use.

### B. GenAI Policies in Syllabi and Computing Education

One prospective avenue for studying how faculty are responding to GenAI is through course syllabus policies (Ali et al. 2025). Prior studies found that many course policies take a negative stance toward AI use, while only a small number present positive views (Moore and Lookadoo 2024). In another analysis of course-level policies, instructors were found to encourage students to document, annotate, and cite their use of GenAI tools (Hooper and Lunn 2025). Similarly, a study found that most instructors defined AI through examples and allowed its use with restrictions (Tong et al. 2025). Research also identified differences between university-level and course-level policies, where university policies focus on broader concerns such as plagiarism and data privacy, while course policies provide more detailed instructions regarding citation, documentation, and classroom activities (Hooper and Lunn 2025). Instructors also prefer guidelines that are clear, comprehensive, flexible, and adaptable to changing AI technologies and diverse student needs (Cacho 2024). Faculty further recognize that GenAI is evolving rapidly and that current rules and guidelines may only have a temporary effect (Mahon et al. 2024).

In computing education, policy stances often vary by academic level and the nature of the programming task. A longitudinal analysis observed a significant rise in GenAI policy adoption within computing syllabi, which increased from 12.8% in early 2024 to 58.8% by late 2025 (Bui and Dong 2026) .Azoulay et al. (2025) described a "colored use" framework (Red, Amber, Green) to delineate where AI is strictly forbidden, permitted for assistance, or fully encouraged. Similarly, Bhalerao (2024) implemented a policy in introductory courses that treats LLMs like human peers - students are encouraged to collaborate and discuss concepts with the model but are prohibited from copying its code outright. Another study showcased an alternative "guardrailed" approach in Harvard's CS50 course, where commercial tools like ChatGPT were limited in favor of a custom-built internal assistant, the "CS50 Duck," designed to guide students toward solutions rather than providing them directly (Liu et al. 2024). While introductory courses have dominated the research landscape, there is an urgent need to adapt upper-level instruction to reflect a workplace where AI proficiency is an expected professional skill (Bouvier et al. 2025). Finally, to ensure long-term academic integrity, Azoulay et al. (2025) advocate for collaborations with AI developers to embed invisible watermarks in AI-generated code, facilitating traceability and encouraging students to produce original work.

### C. Faculty and instructor use of GenAI

Empirical studies show that faculty use and enthusiasm for GenAI significantly lag behind students' use of the technology (Cacho 2024; Dewan et al. 2025). These studies also highlight a general lack of AI knowledge among faculty, including understanding of how AI can be used within their own disciplinary domains and not only for teaching purposes (Kohnke et al. 2023). As a result, faculty who have already adopted GenAI guidelines in their courses can be considered early adopters of GenAI in education. At the same time, faculty express strong concerns about plagiarism and about the potential reduction in students' cognitive development and learning, since it is often unclear what kinds of thinking processes and reasoning are actually being engaged when students use GenAI tools (Khlaif et al. 2024; Lee et al. 2024). This tension is also reflected in how faculty propose to respond to GenAI use in classrooms. For example, some suggest designing AI-resistant assignments, using paper-based exams, incorporating non-text-based assessments, or shifting toward more process- and design-based assignments (Lau and Guo 2023). However, other studies argue that while avoidance strategies may work in the short term, the increasing use of GenAI will require a broader reevaluation of teaching and assessment practices, particularly in computer science education (Prather et al. 2023).

## D. Research Questions

A recent analysis by Hooper and Lunn (2025) suggests that university-wide policies focus more on broader ethical concerns like plagiarism and academic integrity whereas course-level policies focus more on the micro level of impact such as such as citations and writing. Their analysis examined two unrelated datasets that did not link institutions to the course-level policies directly. Therefore, although they found some differences based on their work, a deeper analysis is needed to better understand this relationship in particular through examination of institutions and courses that are related. In this paper we use a related set of data to ask the following research questions: (i) How are institutional guidelines implemented at the Computing course level? (ii) What are aspects of convergence and divergence between institutional and Computing course guidelines?

# III. Research Study

In this paper, we use the intersection of two datasets: (i) institutional-level guidelines and (ii) course-level guidelines from the same R1 universities in the U.S., as designated by the Carnegie Classification. We focus on institutions that have both types of guidelines or policies to compare them. This approach is similar to the one followed by (Brown & Klein, 2020) where they conducted a policy discourse analysis of 151 university policy statements on student information privacy and responsible use of student data from 78 public and private post-secondary institutions in the United States.

## A. Dataset

In the first stage of data collection, the Carnegie list of R1 institutions was used, following Brown & Klein (2020). The list included 131 institutions. The researchers searched each university website for policies or mentions of generative "artificial intelligence" ("AI" or "ChatGPT") in the classroom and used Google searches to confirm no missing information. Only publicly available data was used under Institutional Review Board approval. Data was collected from October 9, 2023, to November 26, 2023. Fourteen institutions had no policy or mention of GenAI, leaving 116 institutions in the analysis. From these, 141 documents related to GenAI policies were collected.

For course syllabi, the search was limited to computer science courses for consistency and because prior work shows CS courses were early adopters of GenAI. Using the same list, Google searches were conducted using "[University name] computer science course syllabus" with ChatGPT, generative AI, and artificial intelligence across spring 2024, fall 2023, and general terms. This resulted in 98 syllabi from 54 R1 universities. Some syllabi (N=6) for the same courses were included as separate documents. Data was collected from March 4, 2024, to May 11, 2024, and each syllabus was treated as a separate unit and coded separately.

## B. Primary Analysis and Findings

For the institutional analysis, all researchers reviewed a random subset of the data (N =20 institutions) and initially discussed codes and developed a codebook. After reviewing another random sample of N =10, this codebook was refined. Two researchers then applied this codebook to another random set of 10 institutions, establishing near perfect agreement to gain inter- rater reliability (IRR). The two researchers coded the remaining dataset separately using a codebook of thirteen codes with corresponding subcodes. Final results are reported out of N=116 and included in Table 1 alongside those for course policy analysis.

For course syllabi analysis one researcher familiar with the process and the data initially reviewed and open-coded the syllabi corpus and created a codebook. Two additional researchers reviewed and coded all 98 syllabi individually using the codebook. The three researchers then met to compare codes and reach agreement working through all 98 syllabi once again. The final codebook consisted of 16 codes. Final results are reported out of N=98. Table 1 presents the primary findings comparatively. The course level guidelines study identified six codes each comprising multiple subcodes.

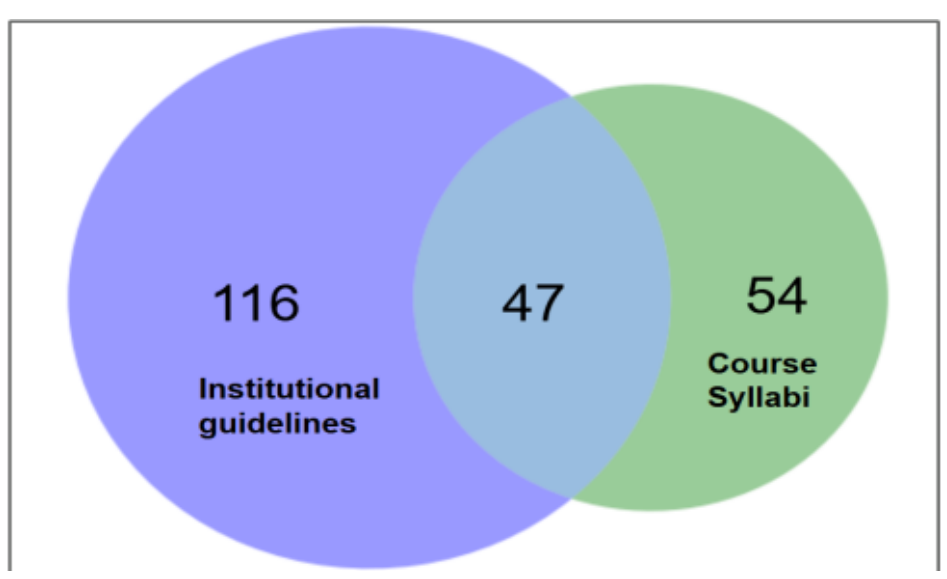

Figure 1: Policy Overlap - Used for Analysis

## C. Secondary Analysis

A secondary analysis was done by two researchers to understand the relationship between the institutional guidelines and course level guidelines. Figure 1 shows among 131 R1 institutes how many institutions have institutional guidelines (116), course level guidelines (54) and how many of them have both (47). We used the same dataset to conduct comparative analysis in three steps. First, we examined the extent to which comprehensive institutional guidelines also provide course level guidelines on the use of GenAI, and the meaning of this alignment. Here, institutions that covered most of the codes can be considered as more comprehensive (Figure 2). Second, we map the institutional-level codes against the course-level codes to identify similarities and gaps in coverage (Table 2). Finally, we present examples from both guidelines for six of the codes to discuss how the similarities and differences are reflected in practice (Table 3).

# IV. Findings

## A. Overlap and Alignment between Institutional and Course-level Guidelines

The prior work related to institutional guidelines defined and captured 13 codes in total. The maximum coverage observed in a single guidelines is 11 codes. Based on the number of codes, institutions were categorized into three groups: comprehensive (those covering 8-11 codes); moderately comprehensive (covering 4-7 codes), and less comprehensive (covering 0-3 codes). Figure 2 shows the distribution of institutions according

Table 1: Major themes and codes captured in the primary analysis and frequency of occurrence

<table>
<tr><th>Code</th><th>Institutional</th><th>Course</th></tr>
<tr><td>Range of Consent</td><td>More than half stipulated syllabus statements for a range of use (N =64, 55%), often in three categories such as “embrace,” “limit,” or “prohibit” (N =43, 37%) vs those that provided more narrow or broader use guidance (e.g., “permissible” and “prohibited” or "no use,” “some use,” “unlimited use,” and “required use") (N =21, 18%).</td><td>Range of use guidelines. Almost all (92%, N=90) syllabi provided explicit guidelines about permissions to use GenAI in their course, with half (50%, N=49) outright prohibiting use and a few only allowing it with explicit permission (N=6). About 41% (N=40) permitted partial use for activities specified in the syllabus.</td></tr>
<tr><td>Encourage/ Discourage</td><td>A majority of universities (N = 73, 63%) encourage the use of GenAI, with many offering detailed guidance for its use in the classroom (N = 48, 41%). More than a quarter discouraged use (N =31, 27%), though some while providing little guidance at all (N =9, 7%). Some provided no sample curriculum or guidance but did not seem to discourage use either (N =12, 10%).</td><td>Beyond having a policy, few syllabi went further to communicate explicit encouragement and/or discouragement (17%, N=17). Syllabi were coded as explicitly Discourages Use of GenAI Tools (11%, N=11) and/or Encourages Use of GenAI Tools (7%, N=7)</td></tr>
<tr><td>Acknowledgement of Use</td><td>Students are encouraged and/or required to not only use GenAI in various ways but also to describe the nature of their use of it, providing details about their queries and, above all, cite it–more than a third (N = 44, 38%) provide formal citation guidelines, which are most often references to the APA style guidelines.</td><td>83% (N=81) syllabi provided explicit guidance on how to cite, when to engage with AI, and/or referenced detection tools. A little over two-thirds (N=65) stated that using GenAI without citation was in violation of the honor code and, in some cases, a violation of academic integrity on par with plagiarism. A quarter of syllabi (N=27) stated that students use informal or formal (e.g., APA) citation of GenAI tools. Some (12%, N=12) stipulated providing detailed notation about how it was used and how a problem might have been solved without GenAI. Very few (N=4) warned students that the school provides detection tools that are at the instructor’s disposal and discretion to use.</td></tr>
<tr><td>Domain-specific Guidance</td><td>Half of the institutions mention the use of GenAI in STEM related courses (N =58, 50%) with most mentioning computer science (N =56, 48%) and fewer discussing math or natural sciences. Engineering is only mentioned by N =7 institutions.</td><td>All syllabi were from computer science related courses.</td></tr>
<tr><td>Concerns with Use</td><td>About three in five institutions (N =69, 60%) caution instructors about privacy concerns with GenAI. Almost every institution that talked about privacy concerns, advised instructors to exercise caution about sharing personal or sensitive data with GenAI.</td><td>34% (N=33) of syllabi mentioned implications for use of GenAI including the veracity of GenAI output (20%, N=20), concerns about privacy (19%, N=19), hindering learning (17%, N=17), and the ability to develop skills (7%, N=7).</td></tr>
<tr><td colspan="3">Aspects exclusively in institutional guidance</td></tr>
<tr><td>Curriculum Guidance</td><td colspan="2">Half suggested that teachers reflect on their approach to teaching and evaluating students (N =58, 50%) leaving it open as to how they approached it for their classroom. Encouraging the use of GenAI for instructor lesson planning was somewhat less common but still discussed by a sizable number of institutions (N =34, 29%). The most common guidance was to use GenAI to design and get feedback about classroom activities and assistance with lesson planning content creation (e.g., creating slides, lecture material, etc.). While a number of institutions provide guidance for curriculum involving GenAI, a good number (N = 31, 27%) also provide curriculum to discourage use.</td></tr>
<tr><td>Ethical Concerns</td><td colspan="2">Slightly more than one half of institutions talked about the ethics of GenAI on a range of topics, including Diversity, Equity and Inclusion (DEI) (N =60, 52%), privacy with GenAI–particularly concerns about entering sensitive data or third-party data sharing among GenAI platforms (N =66, 57%)–and the need to have discussions with students about the ethics of using GenAI in the classroom (N =61, 53%).</td></tr>
<tr><td colspan="3">Aspects exclusively in course policies</td></tr>
<tr><td>Anthropomorphism of GenAI</td><td colspan="2">39% (N=38) syllabi were coded as Anthropomorphism of GenAI meaning they attributed living characteristics to GenAI, either as a GenAI Assistant (36%, N=35) or otherwise providing Characterization/ Personification of GenAI (5%, N=5). Below we describe these codes in more depth.</td></tr>
<tr><td>Specific Tools</td><td colspan="2">Nearly all syllabi (N=91) mentioned specific tools; “ChatGPT” (N=90) was mentioned the most, followed by “CoPilot” (N=25). A few mentioned “Bard” (N=12), “Bing AI” (N=7), “CoPilot” (N=4), and “Grammarly” (N=2).</td></tr>
</table>

to code coverage range and indicates percentage of institutions that also provided course level guidelines.

- Comprehensive guidelines. Institutions that have the most comprehensive (8 - 11 codes) GenAI guidelines also include course level guidelines at a moderate to high rate (48.3 - 71.4%). That indicates comprehensive institutional policies do translate better to course level. Comprehensive institutional policies likely include specific guidance/ or encourage faculties in case of implementing course level guidelines.
- Moderately comprehensive guidelines. The institutional guidelines that are moderately comprehensive (4 - 7 codes) offer course-level guidelines at a below average level (30.8 - 31.2%). These universities are likely to be in a transition phase and focused on institutional policy development rather than implementation support to faculties. On the other hand, faculty at these institutions are introducing their own guidelines or waiting for clearer guidance from institutions.
- Less comprehensive guidelines. Less comprehensive guidelines (0 - 3) suggest a bottom-up policy development phase (35.3 - 44.4%). Which indicates faculties are still developing their own guidelines in the absence of comprehensive guidelines from institutions.

*B. Alignment and Divergence in Codes and Subcodes at Institutional and Course Levels*

Table 2. compares the codes and subcodes from both levels. The degree of alignment rated on a scale from high to low - where high mapping level indicates strong similarity of what they intent to capture, medium indicates indirectly capturing similar theme, and low refers less similarity. In addition to the interpretation of the alignments between the codes, institution level gaps and course level gaps have been identified and discussed in this section.

**Institutional level gaps:**

- Course level guidelines mentioned specific GenAI tools, but institution level doesn't.
- There are additional requirements which are more specific in course level guidelines indicating a gap of specifying additional/ exceptional requirements in institutional guidelines.
- In case of anthropomorphism of GenAI, though institution level identified customization capability of GenAI as one of the subcodes it shows less emphasis on this subject compared to course level.

**Course level gaps:**

- The code - 'Syllabus statements' that refers to language that institutions provide to instructors to include in their syllabi from institutional guidelines didn't appear at the course level. This is not a gap as the various codes show the actual implementation of GenAI guidance on the course syllabi.
- The code 'open guidance' from institutional guidance encourages instructors to reflect on what GenAI means for their teaching which is missing from course level implementation because syllabi is for students. However, it is necessary to share the reasoning behind their decisions with students so that they know the rationale behind the rules, not just the end results.
- There are broader social issues such as DEI which are less emphasized in the course level.

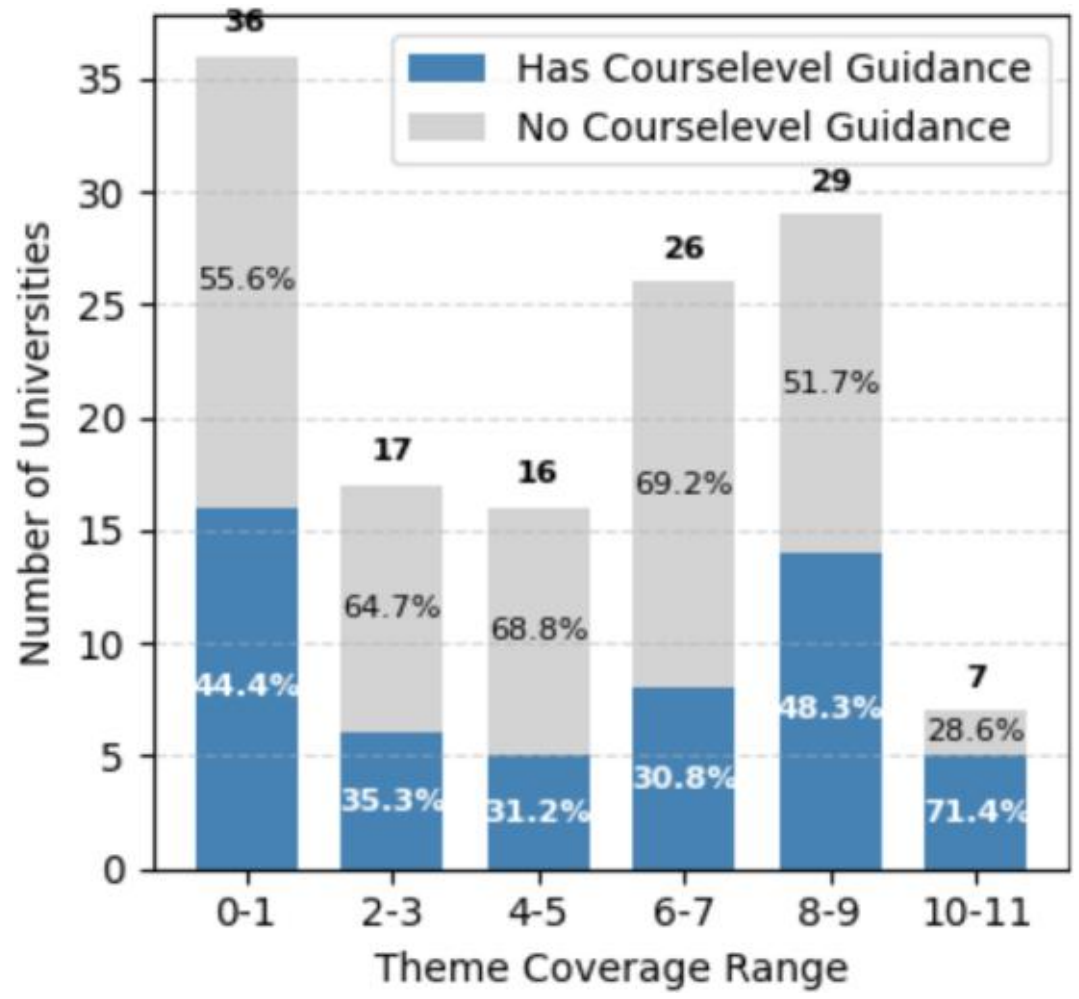

Figure 2: Institution offering guidelines also offering course level guidelines.

## V. Discussion and Implications

Our comparative analysis of institutional and course-level guidance for GenAI shows mixed messages at both levels about whether and how to use GenAI. This uncertainty comes from the fast pace of the technology and changing access controlled by companies developing these tools. As a result, institutions and educators are playing catch up to student use of GenAI. Many elements of institutional guidance are reflected in course policies and course design. At both levels, GenAI use ranges from "allowed" to "prohibit," with variation in how use is encouraged or discouraged. Both levels emphasize transparency and provide guidance on how to acknowledge or cite GenAI use. Both also raise concerns related to data privacy and incorrect information from GenAI tools. Institutional guidance also addresses course and curriculum design and asks instructors to reflect on teaching and assessment in light of this technology. At the course level, GenAI is sometimes described in anthropomorphic terms such as a "tutor" or "coach," reflecting faculty and student mental models of teaching.

In terms of domains, institutional guidance focused largely on STEM disciplines, especially computer science (CS). This is one reason we focused our course policy analysis on CS courses, which provided consistency and allowed us to study early adopters of GenAI in teaching. High uptake in CS is not surprising, as software programming especially at the introductory level is a key use case for GenAI. This is also influenced by training data sources used by GenAI systems, including programming discussions on StackOverflow and Reddit and open-source course repositories online. Most studies in the literature span multiple domains, so future work could examine whether GenAI use differs across disciplines. Beyond

Table 2: Mapping of Institutional-Level and Course-Level Codes showing Degree of Alignment.

| **Codes and subcodes (Course Level)** | **Codes and Subcodes (Institution Level)** | **Mapping Level** | **Interpretation** |
|---|---|---|---|
| Range of Consent, such as prohibited, permitted partially, requires permission | Providing three GenAI use guidelines for instructors (e.g., “embrace”, “limit”, “prohibit”), Provide narrower or broader set of GenAI use guidelines for instructors (e.g. “No use,” “Some use,” “Unlimited use,” “Required use") | High | Both guidelines identified this code by mentioning permissible use, restrictions and partial use. |
| Encourage/ Discourage use of GenAI tools. | Identified under **classroom activities** such as using GenAI for support to deepen discussions, debate with, brainstorm, draft, etc. Teaching GenAI skills (e.g., prompt refinement or “prompt engineering”) | Medium | Course level guidelines are more explicit about encouragement. For example, the code description mentions instructors may limit encouragement (i.e., encourages use as GenAI Assistant but not for other assignments). |
| | Has the code **Discourage GenAI detection tools, Assignments to discourage use, talk with students:** By discussing concerns such as loss of learning opportunities/quality. | High | |
| Acknowledgement of Use / transparency in case of using GenAI tools, citations and disclosure, referencing violation of academic integrity policy and additional requirements. | **Include citation reference format.** | Medium | The importance of acknowledging GenAI use and maintaining transparency - both formally and informally - is directly identified at the course level. While similar themes are captured at the institutional level, they are distributed across multiple codes and subcodes rather than consolidated into a single, focused code. |
| | **Open Guidance:** Encourage instructors to reflect on what GenAI means for their teaching.<br>Codes related to **Discourage GenAI detection tools, Listing GenAI detection tools** also reflect on transparency. | High | |
| | **Privacy:** Caution with sensitive or private information, copyright Caution about legal implications (FERPA, HIPAA, etc),<br>**Talk with students:** Academic integrity and plagiarism, Bias and inaccuracy of GenAI output, Data privacy concerns. | High | |
| | **Classroom activities:** Exploring strengths and weaknesses of GenAI (e.g., through evaluation of veracity, fact checking, writing). | High | |
| Specific Tools, such as mentioning tool’s name (eg. ChatGPT, Claude) | N/A | Low | No code appeared/ have been captured for the institution level guidelines regarding different GenAI tools. |
| Concerns with Use. For example, implications on skill development, learning, privacy and awareness, veracity of GenAI output. | **Classroom activities:** Teaching GenAI skills<br>**STEM related use, Talk with students:** Intellectual - based skill atrophy (e.g., complex problem-solving) | High | Both the institution level and course level guidelines mention implications through different codes and subcodes. |
| | **Privacy**: Caution with sensitive or private information, copyright.<br>**DEI:** Bias in GenAI output can harm underrepresented or underprivileged students through the usage of microaggressions.<br>**Talk with students:** Academic integrity and plagiarism, Bias and inaccuracy of GenAI output, Data privacy concerns | High | |
| Anthropomorphism of GenAI such as GenAI as tutor and assistant, Characterization/ Personification of GenAI. | **Classroom activities:** Using GenAI for support to deepen discussions, debate with, brainstorm, draft, etc. | Medium | The code from institutional guidance “Lesson planning” mentioned customize learning as one of the subcodes but it is not as emphasized as in course level guidance that says, "Instructors appear to personify GenAI tools by directly comparing them to external entities that have abilities beyond general tools." |
| | **Lesson planning:** Customize learning (Providing custom feedback, curriculum, and learning tools for students) | Medium | |

Table 3: Examples of Guidance from Institutional and Course Policies (number in parentheses refers to institution number in our dataset and the corresponding policy)

| Examples | Institutional-level | Course-level |
|---|---|---|
| **Codes related to range of consent, such as prohibits use, partially prohibits and permissible use.** | "[University] will not prescribe a formal policy for the use of AI, rather the importance of empowering instructors to make informed decisions based on their pedagogical goals, subject matter, and student needs will be followed. Instructors should refrain from sharing or inputting student work into online AI tools, including AI detection tools, without obtaining student consent." [94b] | "Using internet resources to find solutions to homeworks is NOT permitted and will be regarded as plagiarism. This means using online search engines, discussion boards, and/or online communities/chats (such as ChatGPT, Discord, Slack, Chegg, StackExchange, Quora, etc) to find solutions is NOT permitted." [88a] |
| **Codes related to Encourage/ Discourage use of GenAI tools.** | "CET encourages faculty to experiment with using AI Generators as a timesaving tool or a starting point." "Instructors should encourage [students] to explore generative artificial intelligence (AI), using these new tools to create, analyze, and evaluate new concepts and ideas that inspire them to generate their own academic work." [127e] | "That is to say, that you may use AI models such as ChatGPT or Claude 2 to help understand the assignments, …. Submitting assignments completely generated by AI is strictly prohibited and when discovered will be awarded 0 points for the assignment." [118j] |
| **Codes related to acknowledgement of Use / transparency about GenAI tools, detection tools, rules and policies, reflections.** | "Whether you're allowing the use of AI tools for certain assignments or banning them altogether, you and your students will benefit from a clear statement of what role AI should play in the class. The following resources, including two from the […] Center for Teaching and Learning, may benefit you in crafting that Language" [84e] | "..the following are activities that may be considered academically dishonest in other contexts, but are acceptable in the project: Using Generative AI systems (we expand on this in the section below). In general, we will not be policing the attribution of small-scale contributions by outside sources, but it is still important that you include those attributions from a documentation perspective." [80a] |
| **Codes related to use concerns and implications, such as skill development and learning, privacy and integrity, output quality etc.** | "It can be difficult to distinguish the outputs of these tools from human-generated content, so concerns in higher education center on potential risks to academic integrity." "Students may be tempted to utilize generative AI tools to generate essays, reports, or solutions to assignments, thus compromising the principles of fairness and personal academic growth." [34f] | "..use of these tools (AI or any tool) is academic misconduct -- this can result in dismissal from your program…Also remember that AI programs should assist with learning. Such a tool is NOT intended to be a SUBSTITUTE for learning. And you'll see that AI programs make mistakes." [31b] |
| **Codes related to Anthropomorphism of GenAI, such as using as GenAI assistant, Personification of GenAI.** | Absent in institutional level guidelines. | "If you are concerned about academic honesty issues, a good heuristic to use in order to figure out what is or isn't appropriate is to imagine GenAI systems as a helpful but fallible classmate. For example, it is generally appropriate to review lecture notes with a classmate and ask questions about concepts you didn't fully understand. So, asking GenAI about this is likely fine." [80a] |
| **Codes related to accessibility and equity, language barrier, bias in output.** | "Because the software is only as good as information it finds and ingests (remember the principle of GIGO: garbage in, garbage out), it may well create prose that mimics structural bias and racism that is present in its source material." [83b] | Absent in course level guidelines. |

STEM, writing courses are another major area of GenAI use, as they are taken across disciplines and writing is a core component of all fields.

In many ways, the implications of this work are clear. Across HEIs, guidance for GenAI use is now common, and although implementation is mixed, the use and acceptance of GenAI at the institutional level and course level is likely to continue. What is unclear is how this affects teaching and learning practices. For many educators, a key justification for including GenAI is preparing students for the future world of work where GenAI will be widely used, and not allowing its use or not using it in instruction may be a drawback. However, despite CS receiving more institutional attention than most other disciplines suggests that increased institutional focus on CS has not translated into clearer direction for CS instructors, leaving them to navigate these decisions largely on their own. In this paradoxical context, based on our findings and reading of the current literature, we believe that it is important to acknowledge that the transition to productive GenAI use in line with the cognitive development we want to see in our students is going to be an arduous task. For CS educators in particular, it will be a burdensome undertaking as substantial revisions to existing pedagogical practices will be needed. The revision of pedagogical practices in CS is not minor. It challenges the structure of CS1/CS2 courses, auto-graded assignments, lab work, and project-based assessment that have been foundational for decades. There is also an additional danger here that developers of GenAI applications will be able to sell their wares to the administrators as shiny objects that can solve all problems easily and cheaply, and like more education reforms this will not be the case. What then do we need to do?

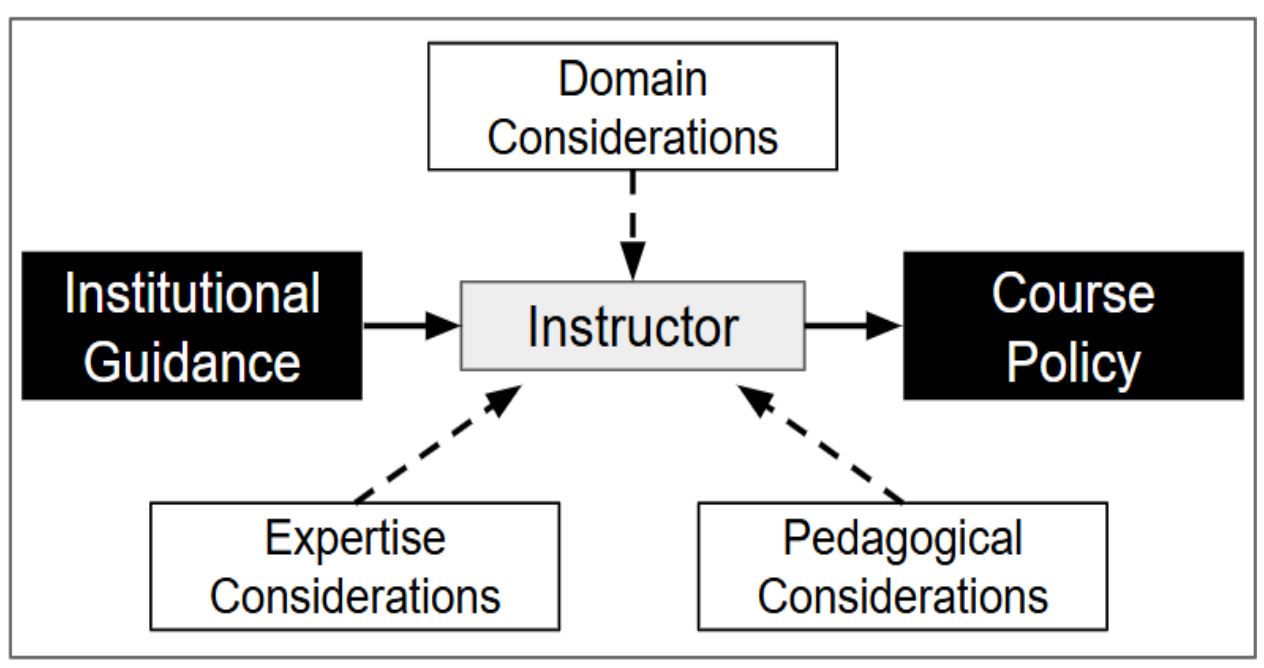

Figure 3: Factors mediating instructor success.

In conclusion, we present a framework that can guide the translation of institutional guidance to course implementation by placing the instructor at the center of the process and considering mediating factors (see Figure 3). The first factor is the instructor's current expertise in GenAI use and broader AI literacy, along with the training needed to implement GenAI in instruction. In CS, AI literacy goes beyond general prompt literacy - it includes understanding model limitations in code generation, recognizing security vulnerabilities in AI-generated code, and knowing when not to trust GenAI output (Cambaz and Zhang 2024)(Fernandez and Cornell 2024).The second factor is discipline or domain. This means accounting for whether the course is theory heavy, such as algorithms or data structures, or applied, such as software engineering or web development, as the implications of GenAI use differ sharply across these contexts. It shapes required expertise, workforce needs, and alignment of course content and GenAI use, though this is often overlooked in institutional guidance and left to instructors. A third factor is the nature of instruction and required pedagogical knowledge, as GenAI use varies across lecture-based, lab-based and project-based courses. Together, these factors shape course policy and determine whether implementation can meet course learning outcomes.

## VI. Limitations

This study has several limitations. First, our dataset is restricted to R1 institutions in the United States as classified by the Carnegie Classification, which limits the generalizability of our findings to other institution types (e.g., liberal arts colleges, community colleges) or to higher education systems outside the U.S., where regulatory environments and institutional cultures around GenAI may differ substantially. Second, our analysis relies on publicly available documents. Institutions or instructors may maintain additional guidance shared only through internal channels (e.g., learning management systems, faculty handbooks, or verbal instruction), meaning our dataset may not fully capture the guidance landscape at each institution. Third, the institutional and course-level data were collected during different time windows (October - November 2023 for institutional guidelines and March - May 2024 for course syllabi). Given the rapid change in GenAI policy during this period, some of the observed divergence between levels may reflect this temporal gap rather than a structural disconnect between institutional and course-level guidance.

Additionally, our course-level analysis focused on computer science syllabi to maintain consistency with prior work and because CS courses are recognized early adopters of GenAI. As a result, our findings about convergence and divergence may not generalize to other disciplines, particularly non-STEM fields where GenAI use cases and instructor concerns may differ. Finally, our analysis is based on the content of written policy documents and does not capture how these policies are actually implemented, interpreted, or enforced in practice. We do not have data on instructor or student experiences, compliance, or the real-world impact of these guidelines on learning outcomes.

## Acknowledgements

This work is partly supported by U.S. NSF Award# 2439459, 2439460, 2319137, 1954556, and USDA/NIFA Award# 2021-67021-35329. Any opinions, findings, and conclusions or recommendations expressed in this material are those of the authors and do not necessarily reflect the views of the funding agencies.